\documentstyle[aps,prl]{revtex}  %% Journal Style

\begin{document}
%\flushbottom
\preprint{UCLA-NT-9801} 
\draft

\twocolumn[\hsize\textwidth\columnwidth\hsize  %**Journal
\csname @twocolumnfalse\endcsname              %**Journal

\title{\bf  Color Mixing in High-Energy Hadron Collisions} 

\author { Chun Wa WONG}

\address{
Department of Physics and Astronomy, University of California, 
Los Angeles, CA 90095-1547}

\date{24 March 1998}

\maketitle

\begin{abstract}
The color mixing of mesons propagating in a nucleus is studied with 
the help of a color-octet Pomeron partner present in the two-gluon 
model of the Pomeron. For a simple model with four meson-nucleon channels, color mixings are found to be absent for pointlike mesons 
and very small for small mesons. These results seem to validate 
the absorption model with two independent color components used in 
recent analyses of the nuclear absorption of $J/\psi$ mesons 
produced in nuclear reactions.
\end{abstract}

%%  Include the following line for Journal style
   ]  %**Journal

\narrowtext

  The two-gluon model of the Pomeron (TGMP) \cite{Low75,Nus75,Gun77,Won96} gives a simple if somewhat 
oversimplified \cite{Kur76,Bal78,Nik94} picture of high-energy 
hadron-hadron scatterings. It makes the interesting prediction 
that color-octet (C8) hadrons interact much more strongly 
with nuclei than color-singlet (C1) hadrons \cite{Dol92,Won96}. 
This color dependence of hadron-nucleon cross sections makes it 
possible to obtain information about the C8 fraction of meson 
precursors produced in nuclear reactions, a topic of considerable current interest.

Analyses of the absorption of $J/\psi$ mesons produced in nuclei 
at high energies using an absorption model with both C1 and C8 
precursors propagating independently have recently been made \cite{Won98,Qia97}. 
In one analysis \cite{Won98}, both the C8 fraction and its absorption 
cross section $\sigma_{\rm abs \, 8}$ by a nucleon are fitted to 
the data for an assumed C1 cross section $\sigma_{\rm abs \, 1}$. 
Fitting the 800 GeV data gives $\sigma_{\rm abs \, 8} \approx$ 
15 (20) mb if $\sigma_{\rm abs \, 1}$ is taken to be 0 (3.5) mb. 
The important change is in the fitted C8 fraction, which 
decreases from 0.7 to 0.5 when $\sigma_{\rm abs \, 1}$ takes on 
the nonzero value.
The data at 200 GeV are not as informative, but they seem to 
prefer smaller values of $\sigma_{\rm abs \, 8}$.

In another analysis \cite{Qia97}, the C8 fraction is first 
deduced from other production data, and only $\sigma_{\rm abs \, 8}$ 
is fitted. A value of 11 mb is obtained by fitting data at 158-450 
GeV, in rough agreement with the more empirical analysis of \cite{Won98}. However, as first pointed out by C.Y. Wong 
\cite{Won97}, the fitted cross sections disagree seriously with 
the theoretical C8 total cross section, which in the TGMP is about 
50 mb \cite{Won96,Dol92}.

Before discussing this descrepancy between theory and experiment, 
we shall first review an older picture of absorption based on a 
single precursor component \cite{Ger88}. The absorption data 
require an absorption cross section of about 7 mb \cite{Ger88,Kha97}, whatever the nature of the precursor. This cross section is 
somewhat larger than that expected of $J/\psi$ mesons of normal size. However, the major problem is that the $c\bar c$ pair originally produced at a point may not have sufficient time to grow to 
full size before it interacts with another nucleon \cite{Kha96}. 
It has been proposed instead that the $c\bar c$ 
pair picks up a hard gluon soon after production and propagates in 
the nucleus as a $[(c\bar c)_8g]_1$ hybrid \cite{Kha96}. The 
hybrid cross section in the TGMP is known to be about 9/4 that of 
the C1 meson with the same mean square radius (between the gluon 
and the pair in the case of the hybrid) \cite{Kha96,Liu94}. 

This hybrid picture is 
very attractive because the hybrid size can be chosen to fit any 
cross section. In fact, it has no trouble accomodating the 
two-precursor model where the unexpectedly small empirical value 
found for $\sigma_{\rm abs \, 8}$ can simply be attributed to 
a precursor produced as a C8 pair but propagating as a hybrid of 
the right size. The hybrid model also addresses the conceptual 
question as to how the C8 pair could remain a free unfettered 
object as it propagates in the nucleus. However, it is fair to 
say that this very promising hybrid picture has not been 
established beyond reasonable doubt.

Let us now concentrate on models in which both C1 and C8 meson precursors are produced. If these precursors are not coupled 
by subsequent interactions with nucleons, they would propagate independently after production. This is the picture assumed in the 
two-precursor models used recently in data analysis 
\cite{Won98,Qia97}. The main purpose of this note is to determined 
if the assumption of independent propagation is justified. This is 
done by studying a model in which the C1-C8 mixing of these meson precursors is induced by a C8 partner of the usual C1 Pomeron. 
The coupled-channel problem is cast in the approximate form used 
in \cite{Huf96,Won96}, and solved for four chosen meson-nucleon channels. The resulting color mixings are found to be absent for pointlike mesons, and are very weak for small mesons: The cross 
sections are practically unchanged, while the predominantly C1 
meson eigen channel has only 1\% admixture of C8 states. These 
results seem to justify the assumption of independent propagation 
of the produced color precursors.

Let us begin by noting that although the C8 and C1 meson states can 
be connected by the exchange of single gluons, such Born amplitudes 
are real and do not contribute to the imaginary part of the forward scattering amplitude appearing in the optical theorem. This means 
that it is necessary to use the TGMP also for the cross-channel 
terms that cause color mixing in the interacting hadrons. This 
involves the exchange of a C8 version of the Pomeron.

To obtain these matrix elements, we simply use the same color 
operator of the TGMP with different channel wave functions. Only 
four different color channels are included in the present study: 
two overall C1 channels $|m_1N_1\rangle$ 
and $|(m_8N_8)_1\rangle$, and two overall C8 channels 
$|m_1N_8\rangle$ and $|m_8N_1\rangle$, where $m_i$ ($N_i$) denotes
a meson (nucleon) state of color C1 (C8) for $i=1$ (8). These 
states will be labeled 1 to 4, in the order presented.

The overall C1 channels do not mix with the overall C8 channels. 
The total cross section is thus described by a $4\times 4$ 
matrix $\Sigma$, with the following nonzero matrix elements:

\begin{eqnarray}
\Sigma_{11} = \sigma(1,1) - \sigma(2,1) - \sigma(1,2) + 
\sigma(2,2),
\end{eqnarray}

\begin{eqnarray}
\Sigma_{12} = \Sigma_{21} = {5\over 8}\sqrt{2}\Sigma_{11},
\end{eqnarray}

\begin{eqnarray}
\Sigma_{22} = {69\over 16}\{\sigma(1,1) - {4\over 23}[\sigma(2,1) + \sigma(1,2)] + \sigma(2,2)\},
\end{eqnarray}

\begin{eqnarray}
\Sigma_{33} = \sigma(1,1) - \sigma(2,1) + 
{1\over 8}[\sigma(1,2) - \sigma(2,2)],
\end{eqnarray}

\begin{eqnarray}
\Sigma_{44} = \sigma(1,1) + {1\over 8}\sigma(2,1) - 
 [\sigma(1,2) + {1\over 8} \sigma(2,2)],
\end{eqnarray}
and
 
\begin{eqnarray}
\Sigma_{34} = \Sigma_{43} = {5\over 16}\Sigma_{11}.
\label{sigma34}
\end{eqnarray}
Here $\sigma(i,j)$ is the absolute value of the contribution when 
the number of quarks involved on the meson (nucleon) side of the interaction is $i(j)$. Thus

\begin{eqnarray}
\sigma(2,2)=8 n_m n_N \alpha_s^2 
\int d^2{\bf k} D^2(k) f_m(4k^2) f_N(3k^2)\,,
\end{eqnarray}
involves both meson and nucleon wave-function form factors, $f_m$ 
and $f_N$, respectively \cite{Low75,Nus75,Gun77,Lev81,Lan87,Dol92,Duc93,Won96}. In this expression, $n_i$ is the number of quarks in hadron $i$, $\alpha_s$ 
is the strong interaction coupling constant, and $D(k)$ is the nonperturbative and nonsingular gluon propagator commonly used in the TGMP. On the other hand,

\begin{eqnarray}
\sigma(1,1)=8 n_m n_N \alpha_s^2 
\int d^2{\bf k} D^2(k)\,
\end{eqnarray}
contains no wave-function form factor at all, while $\sigma(2,1)$ involves only the meson form factor, and $\sigma(1,2)$ involves 
only the nucleon form factor. Equation (\ref{sigma34}) is greater 
than the result reportedf in \cite{Dol92} by a factor of 5/2.

For pointlike mesons, $\sigma(2,j) = \sigma(1,j)$. $\Sigma_{11}$ 
and all the channel-coupling terms then vanish. The C1 and C8 
meson precursors then propagate independently of each other.

The color mixing does not vanish for mesons of finite size and 
therefore nonzero mixing matrix elements. The question is also interesting because of a minor complication: The mesons (including 
the C8 ones) see a C1 nucleon on approach, but leave it behind 
after scattering partially in C8 states. This means that 
the nucleons are not in color eigenmodes except in the trivial 
non-coupling limit.

With C1 (C8) mesons approaching in channels 1 (4) and exiting in 
channels 1 and 3 (4 and 2) at each nucleon site, one can construct 
a simple mixing problem with 2 meson channels for meson-nucleus 
scattering if one assumes that each pair of meson exit 
amplitudes add coherently. The mixing is then described 
by the $2\times 2$ matrix

\begin{eqnarray}
\left( \begin{array}{cc}
       \Sigma_{11} & \Sigma_{34} \\
       \Sigma_{21} & \Sigma_{44} 
       \end{array}\right) 
= \Sigma_{11} \left( 
\begin{array}{cc}
1 & {5\over 16} \\
{5\over 8}\sqrt{2} & x  
\end{array}\right),
\end{eqnarray}
where

\begin{eqnarray}
x = {\Sigma_{44}\over \Sigma_{11}} \approx 
{48\, {\rm mb}\over 5.7\, {\rm mb}} = 8.42.
\label{numbers}
\end{eqnarray}
This matrix is real, but not symmetric. It is diagonalized by a similarity transformation that is not unitary.

When expressed in units of $\Sigma_{11}$, only the matrix element 
$x$ depends on the meson size. Using the numerical values from \cite{Won96} shown in Eq. (\ref {numbers}), the eigenvalues are 
found to be 0.96 $\Sigma_{11}$ and 8.46 $\Sigma_{11}$, while the corresponding eigenvectors are (0.993, -0.118) and (0.042, 0.999).
These color mixings are very weak; the mixing is stronger in 
the first state where the C8 fraction is only 1\%. 

Two conclusions could be drawn from these results. First, the meson
precursors of different color structures, once produced, do appear 
to propagate quite independently of each other. Secondly, the weak 
color mixing found in our model does not match the picture sketched 
in \cite{Won97} of 10-20\% admixture of C8 components in a single ``coherent'' meson precursor propagating in the nucleus with a 
single absorption cross section.

I would like to thank Dr. Cheuk-Yin Wong for helpful discussions.

\end{document}